# Unraveling the Metastability of $C_n^{2+}$ (n=2−4) Clusters

Zirong Peng,[†] David Zanuttini,[‡,§] Benoit Gervais,*[,§] Emmanuelle Jacquet,[§] Ivan Blum,[‡]

Pyuck-Pa Choi,[†,∥] Dierk Raabe,[†] Francois Vurpillot,[‡] and Baptiste Gault*[,†]

[†]Department of Microstructure Physics and Alloy Design, Max-Planck-Institut für Eisenforschung GmbH, Max-Planck-Straße 1, 40237 Düsseldorf, Germany

[‡]Normandie Univ, UNIROUEN, INSA Rouen, CNRS, GPM, 76000 Rouen, France

[§]Normandie Univ, ENSICAEN, UNICAEN, CEA, CNRS, CIMAP, 14000 Caen, France

[∥]Department of Materials Science and Engineering, Korea Advanced Institute of Science and Technology (KAIST), 291 Daehak-ro, Yuseong-gu, Daejeon 305-338, Republic of Korea

**Corresponding Authors**

*Email: gervais@ganil.fr (Benoit Gervais).

*Email: b.gault@mpie.de (Baptiste Gault).

**Abstract**

Pure carbon clusters have received considerable attention for a long time. However, fundamental questions such as what the smallest stable carbon cluster dication is remain unclear. Here, we investigated the stability and fragmentation behavior of $C_n^{2+}$ (n=2−4) dications using state-of-the-art atom probe tomography. These small doubly charged carbon cluster ions were produced by laser-pulsed field evaporation from a tungsten carbide field emitter. Correlation analysis of the fragments detected in coincidence reveals that they only decay to $C_{n-1}^+ + C^+$. During $C_2^{2+} \rightarrow C^+ + C^+$, significant kinetic energy release (~5.75−7.8 eV) is evidenced. Through advanced experimental data processing combined with *ab initio* calculations and simulations, we show that the field evaporated diatomic $^{12}C_2^{2+}$ dications are either in a weakly bound $^3\Pi_u$ and $^3\Sigma_g^-$ state, quickly dissociating under the intense electric field, or in a deeply bound electronic $^5\Sigma_u^-$ state with lifetimes longer than 180 ps.

**TOC Graph:**

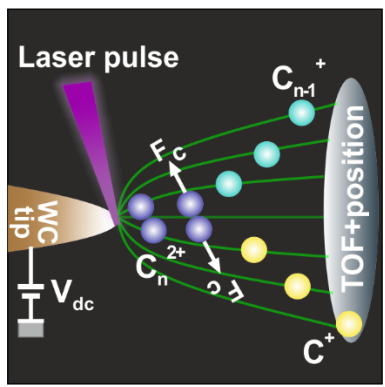

Since first detected in the tail of a comet in 1882,[1] pure carbon clusters have attracted considerable attention from physicists and chemists, which translated into a vast amount of literature.[2–7] Carbon is one of the most abundant elements in the universe. Large to giant carbon clusters such as fullerenes, carbon nanotubes and graphene are materials with rare combinations of mechanical and functional properties. [5,8–10] In turn, small carbon clusters play significant roles in combustion, astrophysics, nuclear and plasma physics.[6,7,11] Besides the importance in numerous technological applications, carbon clusters are also highly valuable to basic theoretical research. C atom can attain three different hybridizations, i.e. sp, $sp^2$ and $sp^3$, which give rise to different types of C−C bonds and clusters with different structures, including linear chains, rings, 2D planes, 3D networks and cages.[2] They provide a unique opportunity to gain deep insights into cluster structure and chemical bonding between atoms.

In 1933, L. Pauling discussed the stability of doubly charged helium dimer ions $He_2^{2+}$,[12] which has opened up a new perspective for understanding bonding and electronic properties of clusters. The stability of a charged cluster reflects the balance between the ionic Coulomb repulsion between the nuclei and the cohesion brought by the electrons. For a specific cluster, the amount of charge it can carry without being torn apart is limited. Similarly, for a particular charge state, only the clusters whose sizes are larger than a so-called *critical size* can resist Coulomb explosion.[13]

Owing to the strong covalent bond and extraordinary stability, fullerene ions $C_{60}^{q+}$ have been studied most extensively among the broad family of carbon clusters.[14] A lot of work has been done to determine the highest charge state that $C_{60}^{q+}$ can reach before it becomes unstable. $C_{60}^{2+}$ and $C_{60}^{3+}$ were observed in the early 1990s.[15,16] Later, intact cations with *q* up to 7 and 10 have been generated using electron impact[17] and ion impact[18,19] for ionization. $C_{60}^{12+}$ produced by infrared radiation[20] is the most highly-charged stable $C_{60}$ cation reported so far. Theoretically, the Coulomb stability limit of $C_{60}^{q+}$ was predicted to be q=18[21,22] or q=14[23–26] using a conducting sphere model or density-functional theory (DFT).

In contrast, information on highly charged small carbon clusters is much scarcer. Fundamental questions such as what the smallest stable carbon cluster dication is remain unclear. Hogreve et al. performed a series of accurate multi-reference configuration interaction (MRCI) calculations on $C_2^{2+}$,[27] $C_3^{2+}$,[28] $C_4^{2+}$,[29] and $C_5^{2+}$.[30] They found that $C_2^{2+}$, $C_3^{2+}$, and $C_4^{2+}$ are all metastable in their lowest electronic state, and $C_5^{2+}$ is the smallest carbon cluster ion exhibiting thermodynamic stability against charge separation.[30] Díaz-Tendero et al. reached the same conclusion by means of DFT and coupled-cluster (CC) theory simulations.[31] Wohrer et al. produced both long-lived and excited $C_5^{2+}$ ions by high velocity collisions of 10 MeV $C_5^+$ ions with a He target.[32] They observed that, conversely to metal cluster ions, whose dominant fission channels are those giving rise to almost symmetric fragments, the $C_5^{2+}$ ions mainly decay to $C_4^+ + C^+$.[33] There are also experimental work reporting the detection of the $C_n^{2+}$ (n=2−4) cluster ions using time-of-flight (TOF) mass spectroscopy,[34,35] but the stability and fragmentation dynamics of the clusters have not been explored in detail.

Here, we focus on the metastability of the three smallest carbon cluster dications, i.e. $C_n^{2+}$ n=(2−4). We conducted a detailed experimental investigation combined with *ab initio* calculations and simulations of the flight of ions from the emitter to the detector. Instead of using the above mentioned conventional ionization methods, we applied laser-assisted field evaporation to produce $C_2^{2+}$, $C_3^{2+}$ and $C_4^{2+}$ dications. Field evaporation is a process whereby the surface atoms of a field emitter, here a sharp needle of tungsten carbide, are desorbed and ionized under the effect of an intense electric field in the range of $10^{10}$ Vm$^{-1}$. Since the cluster ions are formed directly on the surface of the emitter upon desorption rather than by interaction with other energetic particles such as electrons, ions and photons, the only source of their internal energy comes from the emission process itself. This is an advantage of our method.

The experiments were carried out using a straight-flight-path local electrode atom probe (CAMECA LEAP™ 5000 XS). Figure 1 (a) illustrates its basic setup typically used for atom probe tomography (APT) experiments. The field emitter with a tip radius $r$ smaller than 100 nm was held at a high positive voltage $V_{dc}$ in the order of 3-10 kV in an ultrahigh vacuum chamber ($10^{-10}$–$10^{-11}$ mbar). As a result, a strong electrostatic field $F = \frac{V_{dc}}{rK_f}$ is created at the emitter surface. $K_f$ is referred to as the geometrical field factor. It is mainly dependent on the shape of the field emitter and the electrostatic environment.[36,37]. During experiments, $V_{dc}$ was adjusted dynamically according to the current tip radius, so that the strength of the electric field $F$ was always slightly lower than that required for field evaporation. Subsequently, picosecond laser pulses were sent to the apex of the emitter to stimulate and time control the field evaporation process. Laser-pulsed field evaporation is a thermally activated process. After the interaction with the laser beam, the temperature of the emitter apex can increase up to a few hundred Kelvin.[38,39] High temperature field evaporation leads to the appearance of cluster ions.[40] Then, it was rapidly cooled down to the base temperature before the next laser pulse to prevent overheating and uncontrolled field evaporation under the electrostatic field alone. In our experiments, the base temperature of the emitter is set to 60 K. The time interval between the pulses is 4 µs and the cooldown period of the emitter is estimated to be shorter than 6 ns based on the TOF spectrum. On the other hand, although the temperature elevation is relatively large, it only corresponds to a few vibrational numbers for $C_2^{2+}$ and remains small with respect to the ground state dissociation barrier of larger clusters, which is a favorable situation to study the Coulomb instability without possible bias induced by the large amplitude vibrational motion and deformation associated to some excess internal energy. The local electrode (LE), placed ~40 μm ahead of the emitting tip, is grounded to produce a well-defined, confined electric field distribution near the emitter, which drops rapidly towards the LE. By this electric field, the emitted ion is accelerated and projected onto a position-sensitive detector where its TOF can be determined as the time period between the emission of the laser pulse and the moment the ion hits the time-resolved detector. The TOF is then converted into a mass-to-charge state ratio for elemental and isotopic identification. The detector used in the atom probe is a delay-line detector. It consists of a stack of microchannel plates for signal amplification and three delay lines to enhance multi-hits detection capability,[41] i.e. when more than one ion are generated by the same pulse. This multi-hit capability is critically dependent on the processing of the signals[42] and underpins the detection of the fragments from the dissociation of a cluster ion, and thus to determine its complex decay dynamics by coincidence mapping. Upon each ion impacts, the propagation of signals along the delay lines causes a detector 'dead time' and 'dead zone', which can lead to the specific loss of certain ions. Consequently, the multi-hit will be recorded as a single hit or the multiplicity of the hit, i.e. the number of ions included in the multi-hit, will be underestimated. This issue is known as detector pile-up. In another publication,[43] we have closely studied the performance of the applied atom probe instrument on the multi-hit detection for C. The effects of the detector pile-up on this work are not critical.

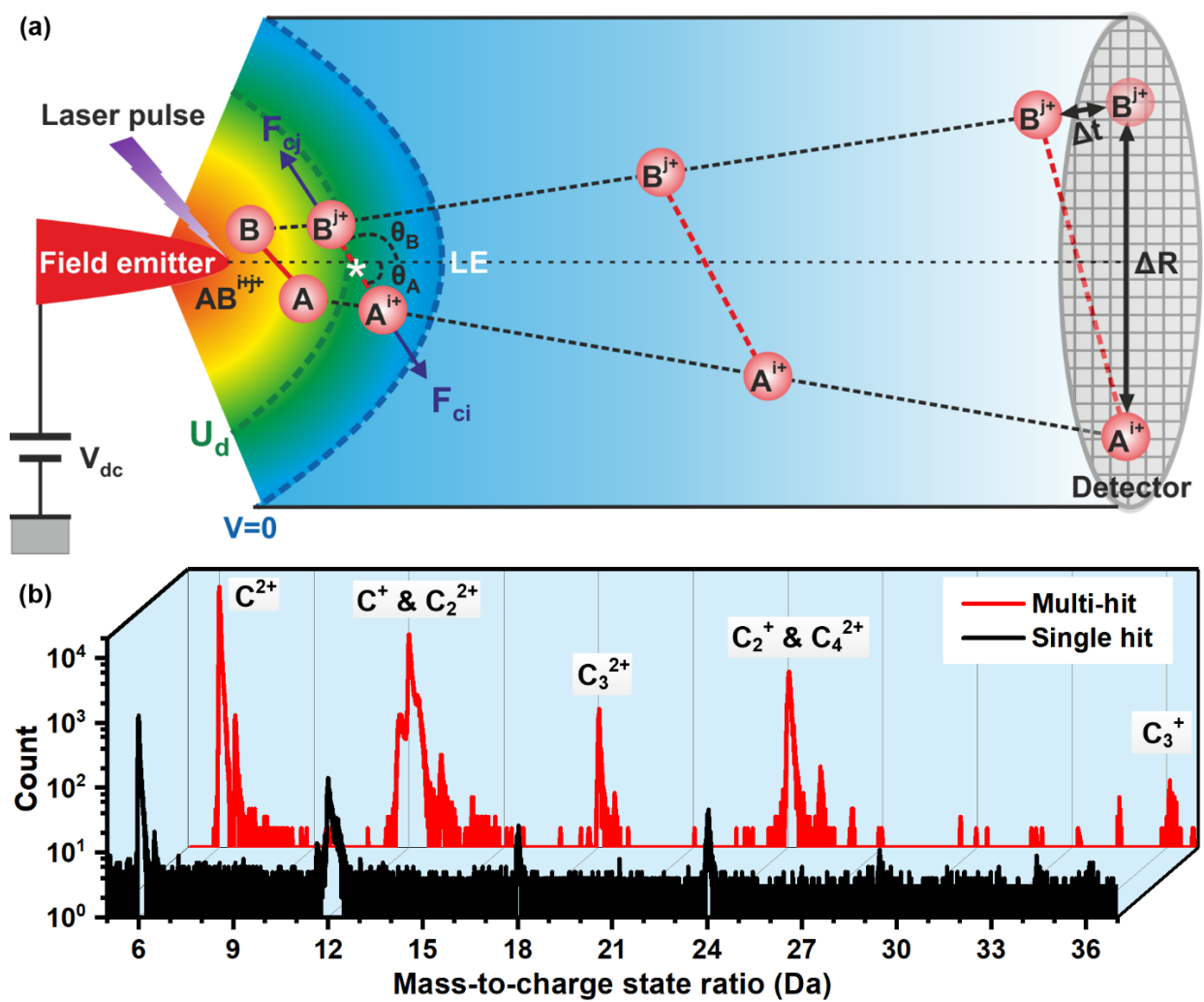

**Figure 1:** (a) Illustration of the experimental setup and a cluster ion fragmentation process. (b) Mass spectrum of carbon ions detected as single hits (black) or multi-hits (red) obtained from a cemented tungsten carbide sample.

Figure 1 (b) shows experimental mass-to-charge state spectra of carbon ions detected as single hits (black) or in multi-hits (red) obtained from a tungsten carbide sample. In this experiment, we detected the multi-hits with a multiplicity up to 10. The relative occurrence frequency decreases with increasing multiplicity. 75.6% of them are double events and 19.9% are triple events. Only less than 1% event with a multiplicity of 9 or 10. The multi-hit spectrum in Figure 1 (b) contains carbon ions from all multi-hits, which are either within the C-C or the C-W type ion pairs. The C-W ion pairs, constituting 93.3%, are mainly formed due to correlated evaporation, i.e. the sympathetic evaporation of adjacent atoms following the evaporation of a first atom.[44] The C-C ion pairs, constituting only 6.7%, are either from correlated evaporation or cluster dissociation that we will discuss in this work. The relative contribution of cluster dissociation is small, but it can be easily separated from the non-dissociative event because of the particular correlation on the mass-to-charge state ratios of the fragments.

Comparing the multi-hit spectrum with the single hit spectrum indicates that most of the carbon ions were detected as part of multi-hit events. This is common for the APT analysis of carbides[35] and is not limited to tungsten carbide. Carbon requires a high electric field for field evaporation, which may cause its retention on the emitter's surface and promote the formation of clusters.[45] Here, we find atomic species, including $C^+$, $C^{2+}$, and cluster species, including $C_2^+$, $C_2^{2+}$, $C_3^+$, $C_3^{2+}$, $C_4^{2+}$. Due to the overlaps of the $C^+$ and $C_2^+$ peaks with the $C_2^{2+}$ and $C_4^{2+}$ peaks, respectively, it is difficult to estimate their relative abundances accurately from the mass spectrum itself, but their existence can be confirmed from the detailed correlation analysis of the multi-hit events. The known natural $^{12}C/^{13}C$ abundance ratio is 92.42.[46] In our measurement, the abundance ratio between the 6 Da and 6.5 Da peak is 90.56, 103.85 and 89.02 for overall, single and multi-hit events mass spectrum, respectively. As we discussed in another publication,[43] some $^{12}C^{2+}$–$^{12}C^{2+}$ multi-hits are incorrectly detected as single $^{12}C^{2+}$ hits when two $^{12}C^{2+}$ ions impacted the detector nearly at the same time and same position. Therefore, the $^{12}C^{2+}$ (6 Da)/$^{13}C^{2+}$ (6.5 Da) ratio measured in multi-hit and overall events is slightly lower than the natural abundance ratio, whereas for single hit events the opposite is observed. Besides, there is no clear indication of the formation of CH type ions. Such species are frequently observed in APT analyses of organic samples,[47,48] but very rare in metal carbides.[35,49] If there is a substantial contribution of $CH^{2+}$ to the 6.5 Da peak, the abundance ratio of 6 Da/6.5 Da peak would be strongly shifted, which is however not the case.

Previously, Liu and Tsong also applied pulsed-laser-stimulated field evaporation and TOF mass spectroscopy to analyze carbon clusters. They observed carbon clusters consisting of up to 11 ions and showing a charge state of up to 3+.[34] Here, we only generated small carbon clusters, arising from the distinctly different experimental conditions. We employed a tungsten carbide field emitter instead of graphite. Furthermore, Liu and Tsong applied a 337-nm-wavelength, 50-µJ-pulse-energy and 300-ps-pulse-width laser pulse, while we utilized a 355-nm-wavelength, 120-pJ-pulse-energy and ~10-ps-pulse width laser pulse. The increase in the temperature of the emitter is expected to be proportional to the peak power, which is five orders of magnitude smaller in our case. In addition, the position-sensitive detector has a good multi-hit capability, which allows us to register information about multiple fragments formed during the dissociation of a molecular ion. The differences in their TOF and hit positions can thus be analyzed to gain insights into the fragmentation process, especially the lifetime of the parent molecular ion.

Metastable cluster ions dissociate during their flight to the detector and give rise to multi-hits. The 12 and 24 Da peaks in the multi-hit mass spectrum are wider than those in the single hit mass spectrum. The $C_3^+$ peak is also more clearly visible. These are indications of cluster ion dissociation. In APT, the measured mass-to-charge state ratio of an ion $m'$ is calculated from its TOF value $t_f$ using the equation $m' \propto 2eV_{dc}\left(\frac{t_f}{L}\right)^2$, where $L$ is the length of the flight path (here ~100 mm). For fragments resulting from in-flight dissociations of cluster ions, their TOF values and thus their measured mass-to-charge state ratios differ from their counterparts directly emitted from the surface. Consider for example the fragment $A^{i+}$ from the dissociation process $AB^{i+j+} \rightarrow A^{i+} + B^{j+}$ sketched in Figure 1. Following the assumption of instantaneous acceleration suggested by Saxey,[50] and taking into account the kinetic energy release (KER) resulting from the fragment repulsion, we can deduce the measured mass-to-charge state ratio $m_A{'}$:

$$\frac{1}{m_A{'}} = \frac{\beta_d}{m_{AB}} + \frac{1-\beta_d}{m_A} + 2\cos\theta_A \sqrt{\frac{\beta_d}{m_{AB}} \frac{\gamma_A}{m_A}} \quad (1)$$

where $m_A$ and $m_{AB}$ are the true mass-to-charge state ratio of $A^{i+}$ and $AB^{i+j+}$ respectively. $\beta_d = \frac{U-U_d}{U}$ ($0 \leq \beta_d \leq 1$) describes the difference between the potential $U$ at the emitter surface and the potential $U_d$ at the position where the fragmentation takes place. The potential $U_d$ is uniquely defined by the lifetime of the parent molecular ion, the applied voltage, and the initial conditions of the parent molecular ion, i.e. its momentum and the position on the specimen's surface. $\theta_A$ is the angle between the Coulomb force $F_{ci}$ and the local electric field generated by the tip. $\gamma_A = \frac{K_A}{2iU}$, with $K_A$ being the KER provided to $A^{i+}$. A similar equation can also be obtained for fragment $B^{j+}$. We provide a detailed derivation of the equation (1) in the supporting information.

The deviations in mass-to-charge state ratio produce dissociation tracks in the multi-hit coincidence map,[51] or more specifically, ion correlation histogram,[50] where the measured mass-to-charge state ratio of a hit is plotted against the mass-to-charge state ratio of another hit within the same multi-hit event. Figure 2 shows the ion correlation histograms for $C_2^{2+}$ (a), $C_3^{2+}$ (b) and $C_4^{2+}$ (c) dications. There is only one particular dissociation track for each species. With the help of equation (1), we can unambiguously identify the dissociation channels from the tracks, which are $C_2^{2+} \rightarrow C^+ + C^+$, $C_3^{2+} \rightarrow C^+ + C_2^+$ and $C_4^{2+} \rightarrow C^+ + C_3^+$ respectively. Previous theoretical work also indicate that the energetically favorable dissociation channel for these small carbon cluster dications $C_n^{2+}$ (n=2−5) is $C_{n-1}^+/C^+$,[27–30] and an earlier experimental study on $C_5^{2+}$ is in good agreement with this theoretical prediction.[33] Irrespective of the low detectability of neutrals in APT, tracks of dissociation channels that result in neutral fragments can also be identified using ion correlation histograms.[52] Here, dissociation channels with neutral fragment emission, i.e. $C_{n-1}^{2+}/C$,

$C_{n-1}/C^{2+}$, and $C_{n-2}^{2+}/C_2$ are not observed.

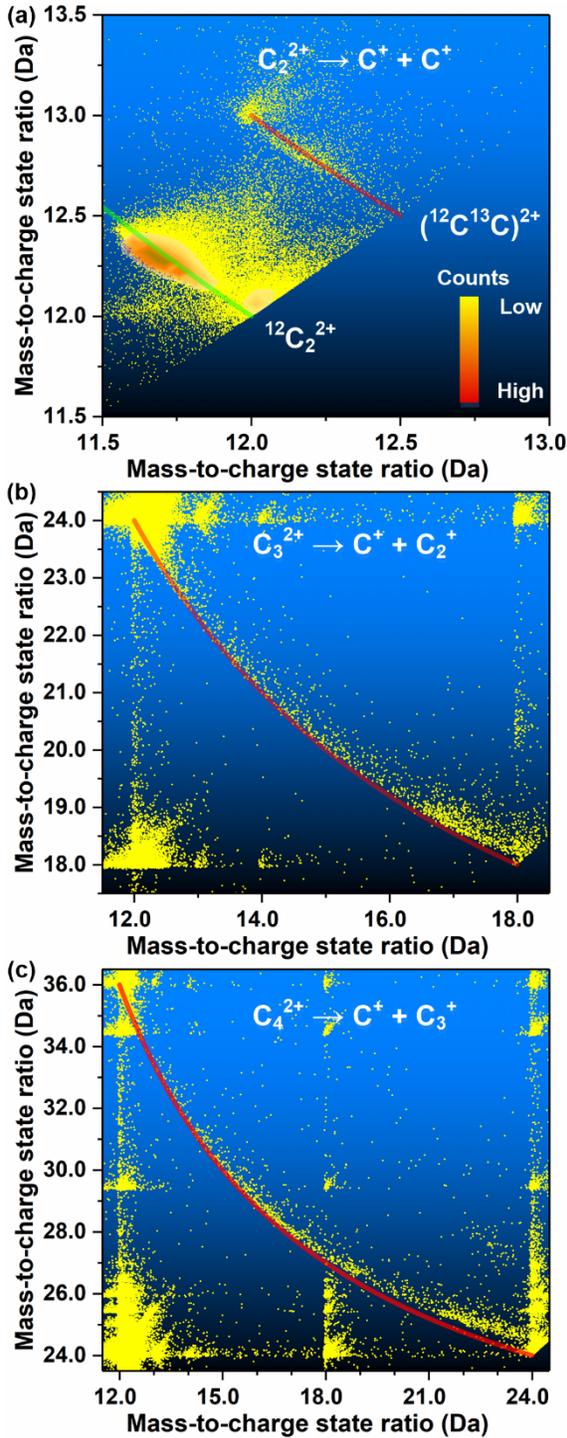

**Figure 2.** Cluster fragmentation tracks observed for (a) $C_2^{2+} \rightarrow C^+ + C^+$, (b) $C_3^{2+} \rightarrow C^+ + C_2^+$, and (c) $C_4^{2+} \rightarrow C^+ + C_3^+$. The green line in (a) corresponds to equation (3) and the red lines correspond to equation (2). In (a), a color map indicating the intensity of the counts is included. The intensity of counts along the track is related to the time probability of dissociation, though not in an unequivocal bijective way.

The red lines in Figure 2 are plotted according to

$$\frac{m_B'}{m_B} = \left(1 - \left(1 - \frac{m_A}{m_A'}\right)\frac{m_B - m_{AB}}{m_A - m_{AB}}\right)^{-1}, \qquad (2)$$

which is just equation (1) without the KER term and after elimination of the parametric $\beta_d$ dependence. According to equation (1), three parameters play a role, namely, the magnitude of the KER itself, i.e. $K_A$, the lifetime of the cluster ion, i.e. $\beta_d$, and the orientation of the cluster ion, i.e. $\theta_A$. With the increase of the cluster ion lifetime, the impact of the KER becomes more and more distinct. For $C_3^{2+} \rightarrow C^+ + C_2^+$, the complete track align well with the red line, which is likely because the cluster axis is perpendicular to the electric field, i.e. $\theta_A = \frac{\pi}{2}$, or the KER itself is small. For $C_4^{2+} \rightarrow C^+ + C_3^+$, we can observe a small offset near the point $(m_{Cn}, m_{Cn})$, i.e. the place where the impact of the KER is expected to be most distinguishable. Both dissociation track of $C_3^{2+} \rightarrow C^+ + C_2^+$ and $C_4^{2+} \rightarrow C^+ + C_3^+$ are long, extending from the point of $(m_{Cn}, m_{Cn})$ to $(m_C, m_{Cn-1})$, but with higher intensity at the $(m_{Cn}, m_{Cn})$ extremity. This means that the majority of $C_3^{2+}$ and $C_4^{2+}$ ions dissociated relatively late, i.e. their characteristic lifetimes are comparable to the TOF range of our spectrometer. The TOF is ~400 ns for $C_3^{2+}$ and ~500 ns for $C_4^{2+}$, depending on the acceleration voltage. As a whole, the proportion of the $C_3^{2+}$ and $C_4^{2+}$ ions that dissociated is large, suggesting that the actual lifetimes of these species are significantly longer than the TOF and beyond the microsecond range. This is consistent with a previous observation of $C_3^{2+}$ ions with TOF longer than 20 μs.[34]

For homolytical dissociation of dications with an even number of carbon atoms such as $C_2^{2+}$ and $C_4^{2+}$, i.e. $C_{2n}^{2+} \rightarrow C_n^+ + C_n^+$, the true mass-to-charge state ratios of the two fragments $C_n^+$ and the parent ion $C_{2n}^{2+}$ are the same, thus the KER is the only cause of a difference in the TOF, which translates into a difference in the measured mass-to-charge state ratio $\Delta m$ between both fragments, giving rise to a dissociation track in the ion correlation histogram. As in this case $\theta_A + \theta_B = 180°$, i.e. $\cos(\theta_A) = -\cos(\theta_B)$, we can eliminate the KER and angle dependence from equation (1) to get:

$$m_A' = \frac{m_A}{2 - \frac{m_B}{m_{B'}}}. \tag{3}$$

In Figure 2 (c), there is no obvious dissociation track corresponding to $C_4^{2+} \rightarrow C_2^+ + C_2^+$, either because the KER is too small to induce noticeable difference in the measured mass-to-charge state ratio of the two $C_2^+$ fragments, or simply because this type of dissociation is not energetically favorable and thus did not occur, as theoretically predicted.[29,31] In contrast, in Figure 2 (a) the dissociation track of $^{12}C_2^{2+} \rightarrow {}^{12}C^+ + {}^{12}C^+$ is very intense, implying a strong KER. The green line in this figure represents equation (3).

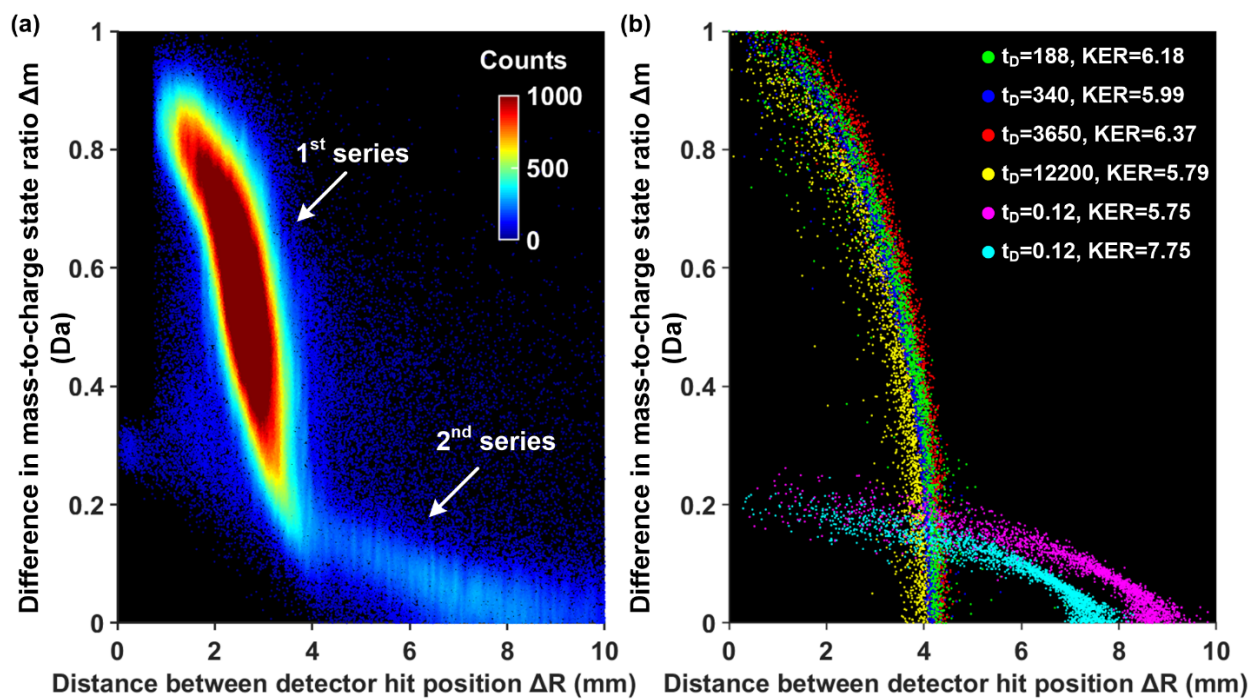

**Figure 3.** Plots showing the relationship between the differences in detector hit position and measured mass-to-charge state ratio of the two $^{12}C^+$ fragments of $^{12}C_2^{2+} \rightarrow {}^{12}C^+ + {}^{12}C^+$, obtained from (a) experiment and (b) simulation. In (b), results of different dissociation time $t_D$, i.e. lifetime of the cluster ions (in ps) and the kinetic energy release KER (in eV) are given. The dissociation distances corresponding to $t_D$, are given in the supporting information.

As Figure 1 (a) illustrates, the KER induced by the repulsive Coulomb forces also causes divergences in the ion trajectories of the fragments ($\Delta R$), which should negatively correlate with the difference in their measured mass-to-charge state ratios ($\Delta m$). The relationship observed for $^{12}C_2^{2+} \rightarrow {}^{12}C^+ + {}^{12}C^+$ is plotted in Figure 3 (a). Surprisingly, we can see two series of events. The first one leads to larger deviations in the measured mass-to-charge state ratio of fragments, while the second one results in larger distances between them. To understand better the dissociation process, we conducted a series of *ab initio* calculations to identify the bound states of $^{12}C_2^{2+}$ that are able to dissociate by means of intersystem crossing due to spin-orbit coupling (SOC). Our calculations to obtain the potential energy curves (PEC) follow closely the MRCI calculation by Hogreve.[27] According to these calculations, $^{12}C_2^{2+}$ has three metastable states, namely $^5\Sigma_u^-$, $^3\Sigma_g^-$ and $^1\Delta_g$. The $^5\Sigma_u^-$ state has a deep potential energy well, with a large number of bound ro-vibrational states. Additional MRCI calculations taking into account a uniform static electric field show that the $^5\Sigma_u^-$ state is the only state able to resist the large electric field in the vicinity of the specimen's surface. Details about the MRCI calculations are summarized in the supporting information. An efficient fragmentation channel for the $^5\Sigma_u^-$ state is the intersystem crossing due to SOC with the $^3\Pi_u$ states. We have computed the lifetimes by means of perturbation theory using the method described in ref.[53] Briefly, the SOC lifetime for a given vibrational level is calculated from the vibrational wave function obtained from the MRCI potential energy curves and using a separate evaluation of the SOC matrix element as described in the supporting information. Subsequently, we simulated the dissociation process of $^{12}C_2^{2+}$ along the $^3\Pi_u$ PEC using an idealized atom probe,[53,54] whose geometry is sketched in Figure 1 (a). Both the tip of the field emitter and the LE are mimicked using a confocal paraboloid following the method given in.[53] The electric field at the tip apex was set at 46 V/nm, which is comparable to the electric fields applied in experiments.[55] The original $^{12}C_2^{2+}$ ions were oriented randomly with respect to the electric field. The distance between the two $^{12}C^+$ fragments at the dissociation time is taken in such a way that the potential energy on the $^3\Pi_u$ PEC equals the sum of the electronic and vibrational energies of the initial state for each vibrational state $v = 0-4$.

Figure 3 (b) shows the relationship between the measured mass-to-charge state ratio difference $\Delta m$ of the two $^{12}C^+$ ions and their distance $\Delta R$ on the detector obtained from the simulations. For the first series of dissociation events, the lifetimes of $^{12}C_2^{2+}$ ions are governed by the inter-system crossing between the $^5\Sigma_u^-$ and $^3\Pi_u$ electronic state. The shortest lifetimes, linked to the lowest-energy vibrational states, are larger than 180 ps and the corresponding KER is larger than 5.75 eV. The dissociation distance ranges from $6\times10^4$ to $4\times10^6$ nm for the dissociation times $t_D$ included in Figure 3. As Figure 3 (b) demonstrates, the exact position of the dissociation track is insensitive to the lifetimes of the cluster ions, but depends on the KER. As the KER increases with the vibrational number $v$, the impact distance $\Delta R$ on the detector progressively increases.

The second series of dissociation events observed in Figure 3 (a) is well reproduced by a dissociation of the $^3\Sigma_g^-$ towards to the $^3\Pi_u$ states, or a direct dissociation of the $^3\Pi_u$ state, with a distribution of KER ranging from 5.75 to 7.75 eV. The ground state $^3\Sigma_g^-$ has a long radiative lifetime by coupling with the $^3\Pi_u$ state.[27] However, the presence of the large electric field near the emitter's surface couples efficiently the $^3\Sigma_g^-$ state with the $^3\Pi_u$ state, leading to a quick dissociation in a time smaller than 1 ps. According to the post-ionization mechanism proposed by Kingham,[56] the $^{12}C_2^+$ ions are supposed to leave the surface of the emitter in their ground state $^4\Sigma_g^-$, and then are ionized by electron tunneling to form one of the three lowest electronic states, i.e. $^5\Sigma_u^-$, $^3\Pi_u$ and $^3\Sigma_g^-$. If the probabilities of removing electrons from different orbitals were the same, then statistically, the abundance ratio would be 5:6:3 for $^5\Sigma_u^-$:$^3\Pi_u$:$^3\Sigma_g^-$. This purely statistical ratio is not observed here, which may be due to the influence of the detector pile-up.[43] When the mass-to-charge state ratio difference of the successive $^{12}C_2^+$ ions in a multi-hit event is smaller than ~0.3 Da, which is exactly the case of the second dissociation event, the detector will very likely only detect one of them. In addition, based on Kingham's work,[56] the critical distance for ionization, where the post-ionization probability peaks, can be calculated. Using the work function of WC (3.6 eV)[57] and the ionization energy of $C_2^+$ (22.48 eV),[31] Kingham's theory predicts that all the $C_2^+$ will be post-ionized into $C_2^{2+}$ at an electrostatic field above 14 Vnm$^{-1}$. The time necessary to reach the critical distance at this field is in the range of 0.25 ps, and could be considered as the lifetime of the $C_2^+$ ion. This result is remarkably consistent with the simulation hypothesis.

In summary, we demonstrated for the first time the metastability of the three smallest carbon cluster dications $C_n^{2+}$ (n=2−4) from both experiments as well as theoretical calculations and simulations. We evidenced that the energetically favorable dissociation channel for them is $C_{n-1}^+/C^+$. For the $C_3^{2+}$ and $C_4^{2+}$ species, we cannot obtain accurate lifetimes values from our experiments. The relatively large proportions of the non-dissociated clusters suggest that their lifetimes extend beyond the microsecond range. For the $C_2^{2+}$ ions, the experimental observation reveals that the tightly bound $^5\Sigma_u^-$ and the quickly dissociating $^3\Pi_u$ and $^3\Sigma_g^-$ electronic states are populated. The lifetime of the metastable $^5\Sigma_u^-$ state ranges from 0.2 to 12 ns, depending on the vibrational state. The distinctly different lifetimes of the $^5\Sigma_u^-$, on the one hand, and of the $^3\Pi_u$ and $^3\Sigma_g^-$ states, on the other hand, lead to different degrees of divergence in their TOF and trajectories during APT analysis. These results support the post-ionization theory proposed by Kingham. The analysis reported here is not limited to tungsten carbide sample and small carbon cluster ions. Laser-assisted field evaporation can provide ready access to diverse types of molecular cations, which can already be noticed from literature. For instance, in the analysis of oxides,[58–62] carbides,[35,43,55] and nitrides,[50,52–54,63] oxygen, carbon, nitrogen as well as (non)metal-oxygen, (non)metal-carbon, (non)metal-nitrogen cluster cations were observed. Large organic molecular ions were detected when measuring biological materials.[47,48] Combining this distinct strength with the time-resolved, position-sensitive detector equipped in APT opens a powerful way to study the stability and complex decay dynamics of multiply charged polyatomic ions.

**ACKNOWLEDGMENTS**

The authors are grateful to Uwe Tezins and Andreas Sturm for their support to the APT, SEM and FIB facilities at Max-Planck-Institut für Eisenforschung GmbH. ZP and BG are grateful to Dr. Michael Ashton from the University of Florida, Gainesville (now in MPIE) and Prof. Susan Sinnott from Penn State University for preliminary discussions and results on the dissociation energies of cluster ions. This work was supported financially by the EMC3 LabEx, AQURATE project and BigMax project (Big-data-driven materials science https://www.bigmax.mpg.de/).

**Supporting Information Available.** 1. Mathematical model of fragmentation tack. 2. Details about the *ab initio* calculation.

**References**

(1) Huggins, W. Preliminary Note on the Photographic Spectrum of Comet b 1881. *Proc. R. Soc. London* **1882**, *33*, 1–3.

(2) Inagaki, M.; Kang, F. *Materials Science and Engineering of Carbon: Fundamentals*, 2nd

ed.; Butterworth-Heinemann: Oxford, 2014.

(3)   Lifshitz, C. Carbon Clusters. *Int. J. Mass Spectrom.* **2000**, *200*, 423–442.

(4)   Varandas, A. J. C.; Rocha, C. M. R. Cn(N=2−4): Current Status. *Philos. Trans. R. Soc. A Math. Phys. Eng. Sci.* **2018**, *376*, 1–46.

(5)   Georgakilas, V.; Perman, J. A.; Tucek, J.; Zboril, R. Broad Family of Carbon Nanoallotropes: Classification, Chemistry, and Applications of Fullerenes, Carbon Dots, Nanotubes, Graphene, Nanodiamonds, and Combined Superstructures. *Chem. Rev.* **2015**, *115*, 4744–4822.

(6)   Van Orden, A.; Saykally, R. J. Small Carbon Clusters: Spectroscopy, Structure, and Energetics. *Chem. Rev.* **1998**, *98*, 2313–2358.

(7)   Weltner, W.; Van Zee, R. J. Carbon Molecules, Ions, and Clusters. *Chem. Rev.* **1989**, *89*, 1713–1747.

(8)   Kroto, H. W.; Heath, J. R.; O'Brien, S. C.; Curl, R. F.; Smalley, R. E. C60: Buckminsterfullerene. *Nature* **1985**, *318*, 162–163.

(9)   Dresselhaus, M. S.; Dresselhaus, G.; Eklund, P. C. *Science of Fullerenes and Carbon Nanotubes*; Academic Press, 1996.

(10)  Thostenson, E. T.; Ren, Z.; Chou, T.-W. Advances in the Science and Technology of Carbon Nanotubes and Their Composites: A Review. *Compos. Sci. Technol.* **2001**, *61*, 1899–1912.

(11)  *Combustion Chemistry*; Gardiner, W. C., Ed.; Springer US: New York, 1984.

(12)  Pauling, L. The Normal State of the Helium Molecule-Ions $He_2^+$ and $He_2^{++}$. *J. Chem. Phys.*

**1933**, *1*, 56–59.

(13) *Clusters of Atoms and Molecules: Theory, Experiment, and Clusters of Atoms*; Haberland, H., Ed.; Springer Series in Chemical Physics; Springer, 1994; Vol. 52.

(14) Echt, O.; Scheier, P.; Märk, T. D. Multiply Charged Clusters. *Comptes Rendus Phys.* **2002**, *3*, 353–364.

(15) Steger, H.; de Vries, J.; Kamke, B.; Kamke, W.; Drewello, T. Direct Double Ionization of $C_{60}$ and $C_{70}$ Fullerenes Using Synchrotron Radiation. *Chem. Phys. Lett.* **1992**, *194*, 452–456.

(16) Böhme, D. K. Fullerene Ion Chemistry: A Journey of Discovery and Achievement. *Philos. Trans. R. Soc. A Math. Phys. Eng. Sci.* **2016**, *374*, 1–16.

(17) Scheier, P.; Märk, T. D. Observation of the Septuply Charged Ion $C_{60}^{7+}$ and Its Metastable Decay into Two Charged Fragments via Superasymmetric Fission. *Phys. Rev. Lett.* **1994**, *73*, 54–57.

(18) Jin, J.; Khemliche, H.; Prior, M. H.; Xie, Z. New Highly Charged Fullerene Ions: Production and Fragmentation by Slow Ion Impact. *Phys. Rev. A* **1996**, *53*, 615–618.

(19) Brenac, A.; Chandezon, F.; Lebius, H.; Pesnelle, A.; Tomita, S.; Huber, B. A. Multifragmentation of Highly Charged $C_{60}$ Ions: Charge States and Fragment Energies. *Phys. Scr.* **1999**, *T80*, 195–196.

(20) Bhardwaj, V. R.; Corkum, P. B.; Rayner, D. M. Internal Laser-Induced Dipole Force at Work in $C_{60}$ Molecule. *Phys. Rev. Lett.* **2003**, *91*, 203004.

(21) Zettergren, H.; Schmidt, H. T.; Cederquist, H.; Jensen, J.; Tomita, S.; Hvelplund, P.; Lebius, H.; Huber, B. A. Static Over-the-Barrier Model for Electron Transfer between Metallic

Spherical Objects. *Phys. Rev. A* **2002**, *66*, 32710.

(22) Zettergren, H.; Jensen, J.; Schmidt, H. T.; Cederquist, H. Electrostatic Model Calculations of Fission Barriers for Fullerene Ions. *Eur. Phys. J. D - At. Mol. Opt. Plasma Phys.* **2004**, *29*, 63–68.

(23) Díaz-Tendero, S.; Alcamí, M.; Martín, F. Coulomb Stability Limit of Highly Charged $C_{60}^{Q+}$ Fullerenes. *Phys. Rev. Lett.* **2005**, *95*, 13401.

(24) Díaz-Tendero, S.; Alcamí, M.; Martín, F. Structure and Electronic Properties of Highly Charged $C_{60}$ and $C_{58}$ Fullerenes. *J. Chem. Phys.* **2005**, *123*, 184306.

(25) Sahnoun, R.; Nakai, K.; Sato, Y.; Kono, H.; Fujimura, Y.; Tanaka, M. Theoretical Investigation of the Stability of Highly Charged $C_{60}$ Molecules Produced with Intense near-Infrared Laser Pulses. *J. Chem. Phys.* **2006**, *125*, 184306.

(26) Sahnoun, R.; Nakai, K.; Sato, Y.; Kono, H.; Fujimura, Y.; Tanaka, M. Stability Limit of Highly Charged $C_{60}$ Cations Produced with an Intense Long-Wavelength Laser Pulse: Calculation of Electronic Structures by DFT and Wavepacket Simulation. *Chem. Phys. Lett.* **2006**, *430*, 167–172.

(27) Hogreve, H. Theoretical Study of the Low-Lying Electronic Spectrum of $C_2^{2+}$. *Chem. Phys.* **1996**, *202*, 63–80.

(28) Hogreve, H. Ab Initio Study of the Dication Carbon Trimer $C_3^{2+}$. *J. Chem. Phys.* **1995**, *102*, 3281–3291.

(29) Hogreve, H. Stability Properties of $C_4^{2+}$. *J. Mol. Struct. THEOCHEM* **2000**, *532*, 81–86.

(30) Hogreve, H.; Jalbout, A. F. The Carbon Pentamer Dication $C_5^{2+}$: Toward Thermochemical Stability. *J. Chem. Phys.* **2003**, *119*, 8849–8853.

(31) Díaz-Tendero, S.; Martín, F.; Alcamí, M. Structure, Dissociation Energies, and Harmonic Frequencies of Small Doubly Charged Carbon Clusters $C_n^{2+}$ (n = 3−9). *J. Phys. Chem. A* **2002**, *106*, 10782–10789.

(32) Wohrer, K.; Chabot, M.; Rozet, J. P.; Gardès, D.; Vernhet, D.; Jacquet, D.; Negra, S. Della; Brunelle, A.; Nectoux, M.; Pautrat, M.; et al. Swift Cluster-Atom Collisions: Experiment and Model Calculations. *J. Phys. B At. Mol. Opt. Phys.* **1996**, *29*, L755–L761.

(33) Chabot, M.; Wohrer, K.; Rozet, J. P.; Gardès, D.; Vernhet, D.; Jacquet, D.; Negra, S. Della; Brunelle, A.; Nectoux, M.; Pautrat, M.; et al. Fragmentation Patterns of Ionized $C_5^{Q+}$ Clusters ( Q = 1 → 4). *Phys. Scr.* **1997**, *T73*, 282–283.

(34) Liu, J.; Tsong, T. T. Kinetic-Energy and Mass Analysis of Carbon Cluster Ions in Pulsed-Laser-Stimulated Field Evaporation. *Phys. Rev. B* **1988**, *38*, 8490–8493.

(35) Thuvander, M.; Weidow, J.; Angseryd, J.; Falk, L. K. L.; Liu, F.; Sonestedt, M.; Stiller, K.; Andrén, H.-O. Quantitative Atom Probe Analysis of Carbides. *Ultramicroscopy* **2011**, *111*, 604–608.

(36) Gomer, R. *Field Emission and Field Ionization*; Harvard University Press: Cambridge, 1961.

(37) Loi, S. T.; Gault, B.; Ringer, S. P.; Larson, D. J.; Geiser, B. P. Electrostatic Simulations of a Local Electrode Atom Probe: The Dependence of Tomographic Reconstruction Parameters on Specimen and Microscope Geometry. *Ultramicroscopy* **2013**, *132*, 107–113.

(38) Kellogg, G. L.; Tsong, T. T. Pulsed-Laser Atom-Probe Field-Ion Microscopy. *J. Appl. Phys.* **1980**, *51*, 1184–1193.

(39) Diercks, D. R.; Gorman, B. P. Nanoscale Measurement of Laser-Induced Temperature Rise and Field Evaporation Effects in CdTe and GaN. *J. Phys. Chem. C* **2015**, *119*, 20623–20631.

(40) Tsong, T. T.; Liou, Y. Cluster-Ion Formation in Pulsed-Laser-Stimulated Field Desorption of Condensed Materials. *Phys. Rev. B* **1985**, *32*, 4340–4357.

(41) Da Costa, G.; Vurpillot, F.; Bostel, A.; Bouet, M.; Deconihout, B. Design of a Delay-Line Position-Sensitive Detector with Improved Performance. *Rev. Sci. Instrum.* **2005**, *76*, 013304.

(42) Da Costa, G.; Wang, H.; Duguay, S.; Bostel, A.; Blavette, D.; Deconihout, B. Advance in Multi-Hit Detection and Quantization in Atom Probe Tomography. *Rev. Sci. Instrum.* **2012**, *83*, 123709.

(43) Peng, Z.; Vurpillot, F.; Choi, P.-P. P.; Li, Y.; Raabe, D.; Gault, B. On the Detection of Multiple Events in Atom Probe Tomography. *Ultramicroscopy* **2018**, *189*, 54–60.

(44) De Geuser, F.; Gault, B.; Bostel, A.; Vurpillot, F. Correlated Field Evaporation as Seen by Atom Probe Tomography. *Surf. Sci.* **2007**, *601*, 536–543.

(45) Gault, B.; Danoix, F.; Hoummada, K.; Mangelinck, D.; Leitner, H. Impact of Directional Walk on Atom Probe Microanalysis. *Ultramicroscopy* **2012**, *113*, 182–191.

(46) de Laeter, J. R.; Böhlke, J. K.; De Bièvre, P.; Hidaka, H.; Peiser, H. S.; Rosman, K. J. R.; Taylor, P. D. P. Atomic Weights of the Elements. Review 2000. *Pure Appl. Chem.* **2003**, *75*, 683–800.

(47) Nishikawa, O.; Taniguchi, M. Scanning Atom-Probe Analysis of Carbon-Based Materials. *19th Int. Vac. Nanoelectron. Conf.* **2006**, 67–68.

(48) Taniguchi, M.; Nishikawa, O. Atomic Level Analysis of Dipeptide Biomolecules by a Scanning Atom Probe. *J. Vac. Sci. Technol. B* **2016**, *34*, 03H109.

(49) Sha, W.; Chang, L.; Smith, G. D. W.; Mittemeijer, E. J. Some Aspects of Atom-Probe

Analysis of Fe-C and Fe-N Systems. *Surf. Sci.* **1992**, *266*, 416–423.

(50) Saxey, D. W. Correlated Ion Analysis and the Interpretation of Atom Probe Mass Spectra. *Ultramicroscopy* **2011**, *111*, 473–479.

(51) Eland, J. H. D. A New Two-Parameter Mass Spectrometry. *Acc. Chem. Res.* **1989**, *22*, 381–387.

(52) Gault, B.; Saxey, D. W.; Ashton, M. W.; Sinnott, S. B.; Chiaramonti, A. N.; Moody, M. P.; Schreiber, D. K. Behavior of Molecules and Molecular Ions near a Field Emitter. *New J. Phys.* **2016**, *18*, 033031.

(53) Zanuttini, D.; Vurpillot, F.; Douady, J.; Jacquet, E.; Anglade, P.-M.; Gervais, B. Dissociation of $GaN^{2+}$ and $AlN^{2+}$ in APT: Electronic Structure and Stability in Strong DC Field. *J. Chem. Phys.* **2018**, *149*, 134310.

(54) Zanuttini, D.; Blum, I.; di Russo, E.; Rigutti, L.; Vurpillot, F.; Douady, J.; Jacquet, E.; Anglade, P.-M.; Gervais, B. Dissociation of $GaN^{2+}$ and $AlN^{2+}$ in APT: Analysis of Experimental Measurements. *J. Chem. Phys.* **2018**, *149*, 134311.

(55) Peng, Z.; Choi, P.-P.; Gault, B.; Raabe, D. Evaluation of Analysis Conditions for Laser-Pulsed Atom Probe Tomography: Example of Cemented Tungsten Carbide. *Microsc. Microanal.* **2017**, *23*, 431–442.

(56) Kingham, D. R. The Post-Ionization of Field Evaporated Ions: A Theoretical Explanation of Multiple Charge States. *Surf. Sci.* **1982**, *116*, 273–301.

(57) Fomenko, V. S. *Handbook of Thermionic Properties: Electronic Work Functions and Richardson Constants of Elements and Compounds*; Samsonov, G. V., Ed.; Plenum Press Data Division: New York, 1966.

(58) Peng, Z.; Rohwerder, M.; Choi, P.-P.; Gault, B.; Meiners, T.; Friedrichs, M.; Kreilkamp, H.; Klocke, F.; Raabe, D. Atomic Diffusion Induced Degradation in Bimetallic Layer Coated Cemented Tungsten Carbide. *Corros. Sci.* **2017**, *120*, 1–13.

(59) Santhanagopalan, D.; Schreiber, D. K.; Perea, D. E.; Martens, R. L.; Janssen, Y.; Khalifah, P.; Meng, Y. S. Effects of Laser Energy and Wavelength on the Analysis of $LiFePO_4$ Using Laser Assisted Atom Probe Tomography. *Ultramicroscopy* **2015**, *148*, 57–66.

(60) Zanuttini, D.; Blum, I.; Rigutti, L.; Vurpillot, F.; Douady, J.; Jacquet, E.; Anglade, P.-M.; Gervais, B. Simulation of Field-Induced Molecular Dissociation in Atom-Probe Tomography: Identification of a Neutral Emission Channel. *Phys. Rev. A* **2017**, *95*, 061401.

(61) Devaraj, A.; Colby, R.; Hess, W. P.; Perea, D. E.; Thevuthasan, S. Role of Photoexcitation and Field Ionization in the Measurement of Accurate Oxide Stoichiometry by Laser-Assisted Atom Probe Tomography. *J. Phys. Chem. Lett.* **2013**, *4*, 993–998.

(62) Blum, I.; Rigutti, L.; Vurpillot, F.; Vella, A.; Gaillard, A.; Deconihout, B. Dissociation Dynamics of Molecular Ions in High Dc Electric Field. *J. Phys. Chem. A* **2016**, *120*, 3654–3662.

(63) Riley, J. R.; Bernal, R. A.; Li, Q.; Espinosa, H. D.; Wang, G. T.; Lauhon, L. J. Atom Probe Tomography of A-Axis GaN Nanowires: Analysis of Nonstoichiometric Evaporation Behavior. *ACS Nano* **2012**, *6*, 3898–3906.